# Orbital Order Triggered Out-of-Plane Ferroelectricity in Magnetic Transition Metal di-halide Monolayers

Xiao-Feng Luo,[1,2,#] Xu He,[3,#], Rui Wang,[4,5,6], Hongjun Xiang[7]*, Jin-Zhu Zhao[1,2,6,8]*

1. Guangdong Provincial Key Laboratory of Quantum Engineering and Quantum Materials, School of Physics, South China Normal University, Guangzhou 510006, P. R. China
2. Guangdong-Hong Kong Joint Laboratory of Quantum Matter, South China Normal University, Guangzhou 510006, P. R. China
3. Theoretical Materials Physics, Q-MAT, CESAM, Université de Liège, B-4000 Liège, Belgium
4. Institute for Structure and Function & Department of Physics & Chongqing Key Laboratory for Strongly Coupled Physics, Chongqing University, Chongqing 400044, P. R. China
5. Center of Quantum Materials and Devices, Chongqing University, Chongqing 400044, P. R. China
6. Center for Computational Science and Engineering, Southern University of Science and Technology, Shenzhen 518055, P. R. China
7. Key Laboratory of Computational Physical Sciences (Ministry of Education), Institute of Computational Physical Sciences, and Department of Physics, Fudan University, Shanghai 200433, China
Shanghai Qi Zhi Institution, Shanghai 200030, China
8. National Laboratory of Solid State Microstructures, Nanjing University, Nanjing 210093, P. R. China.

**Abstract**

Although multiferroics have undergone extensive examination for several decades, the occurrence of ferroelectricity induced by orbital order is only scarcely documented. In this study, we propose the existence of spontaneous ferroelectric states featuring a finite out-of-plane polarization in monolayer compounds of magnetic transition metal di-halides. Our first principles analysis reveals that partially occupied *d*-orbital states within octahedra exhibit a preference for spatial orbital order within a two-dimensional lattice. The absence of inversion symmetry, arising from orbital order, serves as the driving force introducing additional electric polarization along the out-of-plane direction. Unlike previous reported orbital orders arising from metal states in lattice, the non-colinear ones we studied in this work relate to the transition between two insulator states. The resultant asymmetric Jahn-Teller distortions are accompanied as the consequence producing additional ionic polarization. Importantly, our findings indicate that this mechanism is not confined to a specific material but is a shared characteristic among a series of monolayer transition metal magnetic di-halides, proposing an innovative form of intrinsic two-dimensional multiferroic physics.

[#]These authors contributed equally to this work.
Correspondence and requests for materials should be addressed to Hongjun Xiang (hxiang@fudan.edu.cn) and Jin-Zhu Zhao (zhaojz@m.scnu.edu.cn)

Multiferroics [1–4], which concurrently exhibit multiple ferroic orders such as ferroelectricity (FE) [5–9], ferroelasticity [10–14], ferromagnetism (FM) [15–18], and others, have recently garnered considerable attention. Multiple electronic degrees of freedom, such as charge ordering [9,19–21], spin ordering [15], and orbital ordering [21–26], can give rise to ferroelectricity in certain materials. In these cases, the polarization is directly linked to the electronic state, making it an intriguing avenue for tuning the electronic structure through applied electric fields. This heightened interest stems from their intricate multi-order coupling physics and potential applications in practical devices. Particularly over the past decades, driven not only by the emergence of novel physics but also by the miniaturization demands of electronic and spintronic devices, the exploration of coupling mechanisms in two-dimensional (2D) systems capable of inducing ferroelectricity through other type of orders has become one of a central focus in both condensed matter physics and materials science.

Unfortunately, due to the mutually exclusive nature of ferroelectricity and ferromagnetism, the magnetoelectric coupling in materials is not only scarce for 2D materials [27] but also rarely reported at bulk level comparing to conventional ferroelectricity that arises from lattice distortions. The co-existence and the coupling between these two orders typically requires additional mechanism [31–35]. For instance, when considering the coupling between charge order and orbital order, the in-plane ferroelectricity appears in the charged transition-metal halide monolayer ($CrBr_3^{-0.5}$) system, which gained significant attention as the proposed 2D multiferroic materials [9].

However, proposed 2D multiferro candidates rarely exhibit out-of-plane (OOP) component of electronic polarization. The prevailing belief is that, due to strong deplorization field, OOP polarizations in thin films and 2D monolayers driven by electronic origin seems unlikely. As evidenced in previously reported type-II magnetic systems like $TbMnO_3$ [5], $MnI_2$ [15], and $NiI_2$ [16–18], the polarization and atomic displacement resulting from non-centrosymmetric magnetic orders exhibit small amplitudes at their bulk level. The electronic driving force responsible for generating OOP polarization is typically not strong enough to yield substantial OOP polarization in thin film and monolayers [36, 37]. It partially explains the fact that the primary mechanism for most reported OOP ferroelectricity (FE) monolayer materials lies in lattice mode coupling [36, 37] rather than electronic orders. Therefore, exploring novel mechanisms that can produce significant OOP polarization through electronic degrees of freedom remains a significant open question and challenge in the field of FE monolayers.

In this study, we explore the induction of multiferroicity through the coupling of spontaneous orbital order and ferroelectric polarizations in monolayers of magnetic transition metal di-halides (TMDH), specifically $MX_2$ (M = Cr, Mn, Fe, Co, Ni, X = Cl, Br, I), while the sample for some of the members have been experimentally obtained [17,38]. The breaking of inversion symmetry is achieved through spontaneous orbital order of partially occupied $d$ electron states. This orbital order is the primary driving force and results in finite electronic out-of-plane (OOP) polarizations. The consequent lattice distortions then lead to additional ionic OOP polarizations. Such mechanism is not limited in a certain compound but shared in a group of TMDH candidates. In the following sections, we will primarily focus on the results of $FeI_2$ as a prototypical example, while briefly presenting findings for other monolayer compounds of $MX_2$.

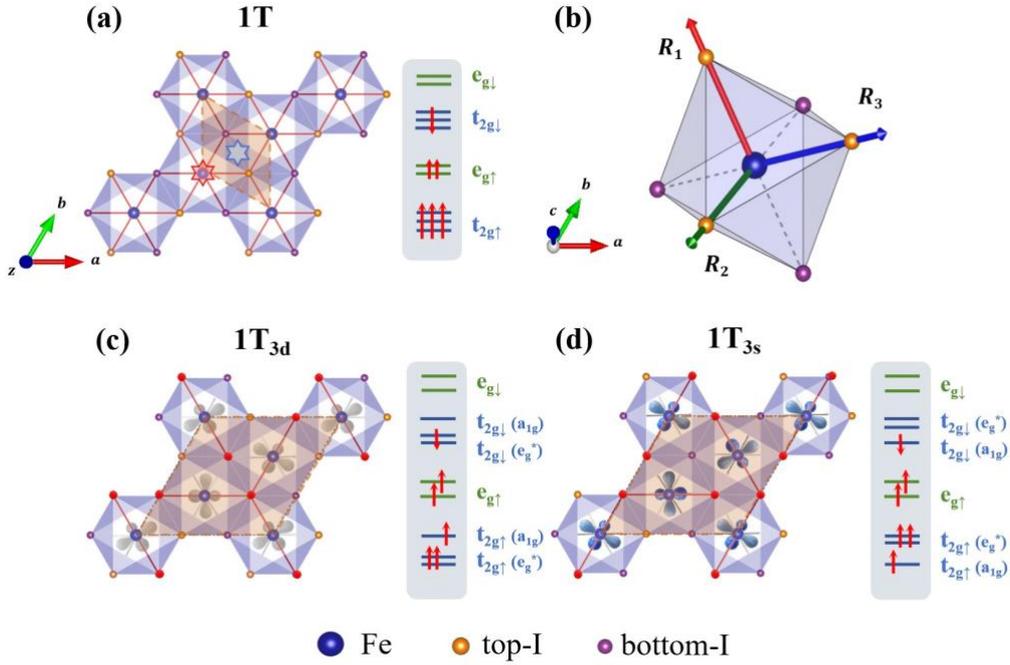

Fig. 1 The top view of illustrated atomic structures and related local reference axis (highlighted by red lines) of octahedrons for (a) 1T phase, (c) $1T_{3d}$ phase, and (d) $1T_{3s}$ phase of monolayer $FeX_2$. The unit cell of each phase is highlighted by dashed lines and light orange color. The Fe atoms are colored by blue while the X (X= Cl, Br, I) atom on the top and bottom surface are colored by yellow and purple, respectively. The illustrated occupation of each $d$-orbitals are shown on the right side of each panel. In (a), the inversion center are highlighted by the red and blue hexagram. (b) single $FeI_6$ octahedron in high symmetric 1T phase where three equivalent local reference axes are shown in red, green and blue colors. In (c) and (d), the reference axis $R$ and the two I atoms along the reference axis are highlighted by red color. In (c) the spatial distribution of empty $t_{2g}$ orbitals of Fe in $1T_{3d}$ phase are illustrated by gray color. In (d) the spatial distribution of occupied $t_{2g}$ orbitals of Fe in $1T_{3s}$ phase are illustrated by blue color.

All parameters of the first-principles calculations carried out in this work are reported in the Section I of our Supplemental Materials (SM). The atomic structure of $FeI_2$ belongs to the hexagonal crystal system of transition metal halide compounds. The highly symmetric phase of the material is the 1T structure, characterized with central inversion symmetry belonging to the $P\bar{3}m1$ layer group (shown in Fig. 1(a)). This structure comprises ABC stacked three atomic layers along the OOP direction in the I-Fe-I sequence. The optimized lattice constant of the p(1x1) formula unit cell is 4.19Å. In this 1T phase, each Fe atom is bonded with six I atoms, forming $FeI_6$ octahedra. Each octahedron has three equivalent reference axes referred to as the $R$ axis. Adjacent octahedrons share a common edge. Due to the triple-folded rotation symmetry of the lattice, three symmetrically equivalent options exist for the local $R$ axis, denoted as $R_1$, $R_2$ and $R_3$, as illustrated in Fig.S1(a) of SM.

The ferromagnetic (FM) state of 1T phase arises from the un-paired $d$ orbital electrons of Fe ions. In the octahedral crystal field forming by $FeI_6$ octahedrons, five $d$ orbitals are divided into two groups which are $e_g$ orbitals containing the $d_{z^2}$ and $d_{x^2-y^2}$ orbitals and t$_{2g}$ orbitals containing $d_{xy}$, $d_{xz}$, and $d_{yz}$ orbitals, respectively. Here, in each individual octahedrons, the $z$ axis of local coordination is defined as the same direction with local reference ***R*** axis while the local $x$, $y$ axis can be defined accordingly. Besides, due to the trigonal crystal fields, the triple degenerate t$_{2g}$ orbitals further split into a single a$_{1g}$ orbital and a double degenerate e$_g$* orbital. As it is shown in Fig. 2(a), the band gap is opened between occupied a$_{1g}$ orbital and empty e$_g$* orbitals showing occupied e$_g$*$_↑^2$a$_{1g↑}^1$e$_{g↑}^2$a$_{1g↓}^1$ configuration, which is refered to as the simplified a$_{1g↓}^1$ for the major spin channel in the following part.

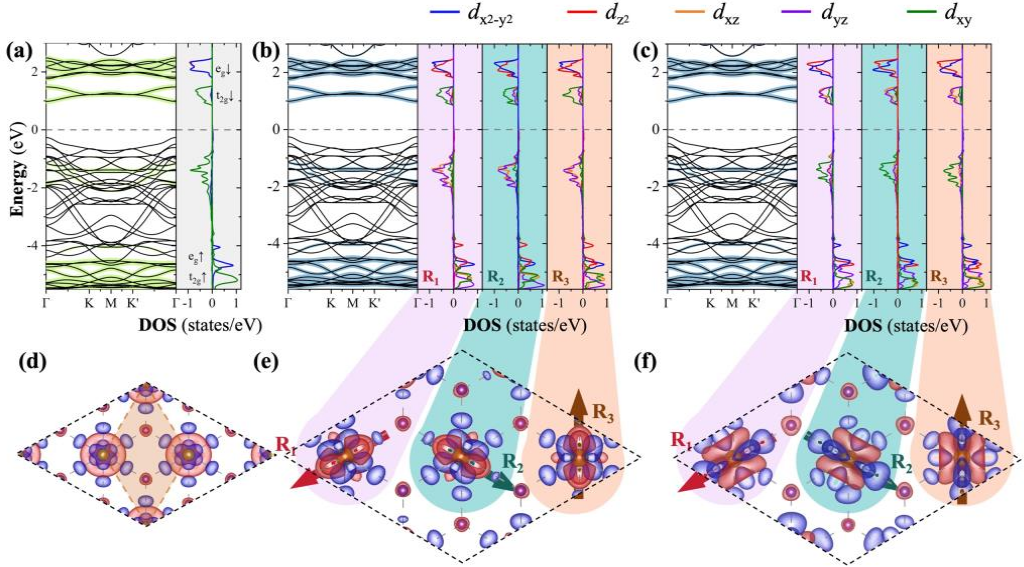

**Fig. 2** The *d*-orbitals projected electronic band structure and density of states (DOS) of FeI$_2$ in (a) 1T phase, (b) 1T$_{3d}$ and (c) 1T$_{3s}$ phase, respectively, where the detailed contribution of DOS from electrons of five *d*-orbitals under the local coordinate of each Fe atom are highlighted by colors. The differential charge density with respect to free atomic reference are shown in (d) for 1T phase, (e) for 1T$_{3d}$ and (f) 1T$_{3s}$ phase, respectively. The region of accumulation and loss of electron are shown by blue and red, respectively. The reference R$_i$ (i=1,2,3) axis of local coordination for each Fe atoms are highlighted in red, green and brown, respectively.

It is worth to mention that, the splitting of *d*-orbitals is defined according to a certain local reference ***R*** axis of octahedron so that the occupied electronic state can be labbelled as the (a$_{1g↓}^1$)$_\mathbf{R}$ state. In the *p*(1x1) high symmetric 1T phase, there are three symmetric equivalent options for the ***R*** axis due to the triple folded rotation symmetry which are R$_1$, R$_2$ and R$_3$ as we shown in Fig. S1. Due to the constrain of the symmetry, three energetical equivalent states, (a$_{1g↓}^1$)$_{\mathbf{R}i(i=1,2,3)}$, need to be considered in 1T phase. As we report in Fig 2 (a) and (d), the electronic states of Fe ions are the mixture of three (a$_{1g↓}^1$)$_{\mathbf{R}i(i=1,2,3)}$ states maintaining inversion symmetry.

However, due to the interaction between occupied a$_{1g}$ orbitals of neighboring FeI$_6$ octahedrons, such high symmetric electronic state of 1T phase is not energetically favored. According to our DFT results, the spatial order of *d* orbital will form spontaneously forming a $p(\sqrt{3} \times \sqrt{3})$ supercell in the lattice by significantly reducing the total energy of 53.33 meV/f.u.. The electronic states are modulated in a larger periodic spatial order, belonging to the polar P3m1 layer group, while all atomic positions are fixed as they are in original *p*(1x1) periodic structures. As shown in Fig. 1(e) and (f), in the modulated supercell, the electronic state of three Fe ions are $(a_{1g\downarrow}^1)_{R1}$, $(a_{1g\downarrow}^1)_{R2}$ and $(a_{1g\downarrow}^1)_{R3}$, respectively while the local main axis of each octahedron varies among $R_1$, $R_2$ and $R_3$.

This ordered state is referred to as the $1T_3$ phase where the triple-folded rotation symmetry is kept. In each FeI$_6$ octahedron, due to the splitting of *d*-orbitals, the two apical I atoms (highlighted in red in Fig. 1 (b) and (c)) are differed with the rest four equatorial ones. Thus, due to the orbital order, the spatial inversion center disappears in $1T_3$ phase which is briefly highlighted in Fig. 1 (e) and (f). During the phase transition from the 1T phase highly symmetric electronic state to the *d*-orbital spatially ordered $1T_3$ phase, the material maintains non-metallic character, which is different from the conventional metal-semiconductor phase transition induced by electronic instability.

The emergence of spatial orbital order in the $1T_3$ phase can be attributed to the partially occupied *d* orbitals of Fe ions. Given that the lattice exhibits a FM state, all electrons near the Fermi level belong to the same spin channel. The magnetic moment of each Fe atom is $3.62\mu_B$ along OOP direction with 1.48eV electronic band gap. Our studies show that, comparing with the rest tested magnetic configurations, the FM state presents the lowest energy (see the section III of SM). Therefore, our following investigation will only focus on the FM state of monolayer FeI$_2$.

The valence electrons prefer hopping between occupied $a_{1g}$ and empty $e_g^*$ orbitals of two neighboring Fe ions as it is illustrated in Fig. S11(b)-(d). When all the octahedra align with the same local ***R*** axis, as in the 1T phase, $\sigma_{d-d}$ bonds form between two partially occupied $a_{1g}$ orbitals of two neighboring Fe ions which gains the total energy of the lattice. In the $1T_3$ phase, where the local ***R*** axis of octahedra varies, $\sigma_{d-d}$ bonds are formed between an occupied $a_{1g}$ orbital and empty $e_g^*$ orbitals or between two empty $e_g^*$ orbitals, representing cases with lower energy. The direction of the local axis for each octahedron varies in the lattice to meet these two requirements, forming the spatial order while maintaining the crystal structures in a highly symmetric configuration. The electronic states of $1T_3$ phase and detailed mechanism of the interaction and electron hopping between occupied and empty orbitals are shown in Section V and Fig. S11 in the Section VI of SM.

Since the $1T_3$ phase is non-central symmetric, finite spontaneous electric polarization is presented with the amplitude of 0.01 μC/cm$^2$ along OOP direction. There are two energetically equivalent states of $1T_3$ phase, named as the $1T_3^+$ and $1T_3^-$ state, corresponding to opposite directions of polarization. These two states are connected by the inversion symmetric operation. The details of symmetry operation connecting $1T_3^+$ and $1T_3^-$ state are illustrated in Fig. S1 of SM.

As mentioned above, the 1T phase of FeI$_2$ exhibits insulating electronic band structures. In the meantime, such insulating feature maintains during the transition from the 1T phase to the 1T$_3$ phase. As shown in Fig. 2, this transition primarily corresponds to changes in orbital occupancy rather than shifts in the energy of the electronic band structures. This mechanism is significantly differed from metal-insulator related phase transition guarantee that external electric field will not be fully screened during the switching of the polarization. Similar features of CrI$_2$ are reported in Sec. V of SM.

As we shown in Fig 2, there are two energetical equivalent electronic state of 1T$_3$ phase, which are refered to as 1T$_{3s}$ and 1T$_{3d}$, respectively. In 1T$_{3s}$ state, the occupation of a$_{1g}$ state is sloley contributed by $d_{xy}$ electrons while it is contributed by $d_{xz}$ and $d_{yz}$ electrons with equal possibility in 1T$_{3d}$ state. These two states will drive different lattices distortion from 1T$_3$ phase that will be discussed in the following part of this work.

Due to the non-central symmetric electronic states of 1T$_3$ phase, additional atomic displacement will be introduced accompanying additional displacive polarizations that can transit the lattice to two local stable phases. The phase with lower energy, with respect to the atomic position in high-symmetric 1T phase, mainly relates to the in-plane trimerization of Fe sublattice and the polar distortion along OOP direction (see the detail in Fig. S1 of SM). Here, we refer to this phase as the Trimer phase showing modulated periodicity of Fe sublattice corresponding to the electronic states in 1T$_{3d}$ phase. The other local stable phase, with higher energy, the Fe sublattice forms an IP non-polar rotation-like motion (see Fig. 3 (c) and (d)). This phase is referred to as the distorted 1T (d1T) phase corresponding to the electronic states in 1T$_{3s}$ phase. The similar structural of this d1T phase is experimentally observed in MoTe$_2$ thin film sample [37,39,40]. Both Trimer and d1T phases belong to the same P3m1 layer group. The potential energy curves (PECs) with respect to the structure evolution of polarization switching between Trimer$^+$, Trimer$^-$, d1T$^+$ and d1T$^-$ are shown in Fig. 3 (e). We can see the 1T$_{3d}^+$ and 1T$_{3s}^+$ state will drive the lattice to Trimer$^+$ and d1T$^+$ phases, respectively. Similarly, the 1T$_{3d}^-$ and 1T$_{3s}^-$ state will drive the lattice to Trimer$^-$ and d1T$^-$ phases, respectively. The stability of each phase have been confirmed by the results shown in Section III and the phonon dispersion curves in Section IV of SM.

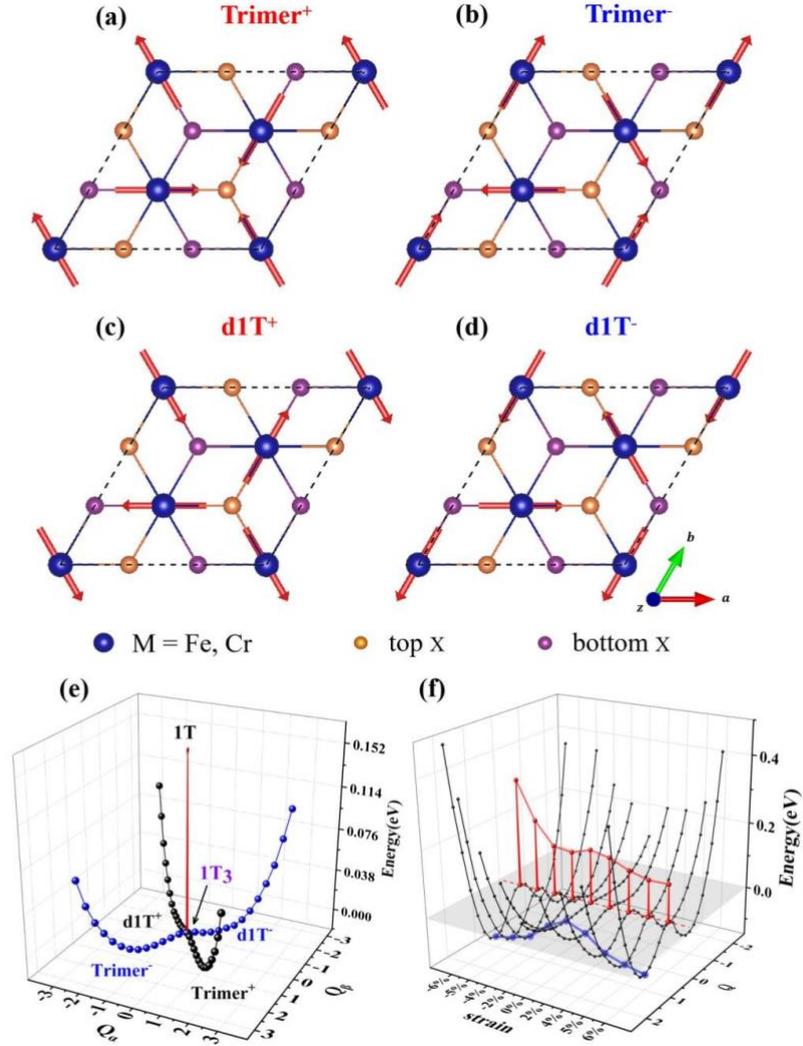

**Fig.3** The in-plane atomic motions, with respect to 1T phase, of (a) Trimer$^+$, (b) Trimer$^-$, (c) d1T$^+$ and (d) d1T$^-$ phases are shown in arrows. The positive and negative signs indicate the direction of the OOP polarizations which are highlighted by red and blue colors. (e) the potential energy curves (PECs) with different polarization directions projected on $Q_\alpha$ and $Q_\beta$ modes which are previous introduced [33]. The red color represents the significant energy reduction resulting from the spontaneous formation of the spatial order of the d-orbitals. The black and blue lines represent the PEC in terms of different OOP, respectively. (f) the PECs of monolayer FeI$_2$ in terms of epitaxial strain where the total energy of the 1T$_3$ phase under each strain are normalized as the reference value zero. The blue curve indicates the local stable states while the red line highlights the evolution of the 1T phase with respect to the epitaxial strain.

Such atomic displacements can be divided into two groups. The first one is the typical Jahn-Teller (JT) type distortions that correspond to the elongation or compression of the octahedron along the local **R** axis, relating to the c-d1T and c-Trimer phase, respectively. Comparing to the reference 1T phase, both the compression and elongation of octahedron along the **R** axis will stabilize the lattice, by 23.5 meV/f.u. and 15.5 meV/f.u., respectively. Notably, the local JT distortion from the splitting of *d*-orbitals in octahedron environment usually preserve the inversion symmetry e.g. in many conventional perovskites. Due to the non-central symmetric orbital order in 1T$_3$ phase, atomic polar distortions are allowed to couple with JT distortions. As it is shown in Fig.S12 of SM, in these two local stable phases, d1T and Trimer phase, the metal ion displaces away from the center of the local octahedron. Additional related data are reported in Section VI of SM (see Fig. S13 of SM).

**Table 1** Corresponding physical quantities, including the OOP polarizations (electronic and ionic contributions) obtained by berry phase approach and the total energy for each studied phase. Polarization values and relative energies in different states are reported, while the energy references are chosen to be the one of 1T$_3$ phase.

|  | 1T | 1T$_3$ | c-Trimer | c-d1T | Trimer | d1T |
|---|---|---|---|---|---|---|
| ele ($\mu C/cm^2$) | 0.000 | 0.012 | 0.041 | 0.075 | 0.051 | 0.090 |
| ion ($\mu C/cm^2$) | 0.000 | 0.000 | 0.000 | 0.000 | 0.133 | -0.031 |
| total ($\mu C/cm^2$) | 0.000 | 0.012 | 0.041 | 0.075 | 0.184 | 0.059 |
| ΔEnergy (meV) | 159.9 | 0.0 | 136.4 | 144.4 | -6.97 | -4.0 |

In the case of FeI$_2$, the Trimer$^+$ phase exhibits a lower total free energy than the d1T$^+$ phase, establishing it as the ground state of the lattice. Both of the two phases present the same direction of OOP polarization, with values of 0.18 $\mu C/cm^2$ and 0.06 $\mu C/cm^2$, respectively. Similarly, the d1T$^-$ and Trimer$^-$ phases show OOP polarizations the values of which are -0.18 $\mu C/cm^2$ and -0.06 $\mu C/cm^2$, respectively. The Trimer$^-$ phase is also energetically degenerated with the Trimer$^+$ phase, while the d1T$^-$ and d1T$^+$ phases share the same total free energy.

Two points need to be noted based on the data in Table I. First, unlike the Trimer phase, the electronic and ionic contributions of OOP polarization in the d1T phase have opposite directions. This aligns with the case of the d1T phase for monolayer MoS$_2$ [37,39,40]. The second relates to the small amplitude of the OOP polarization, both of which are below 0.1 $\mu C/cm^2$. Although the 1T$_3$ phase belongs to a polar group, the asymmetric spatial distribution of the electronic state only provides a relatively small polarization compared to typical ionic polarization, e.g., in monolayer In$_2$Se$_3$ [36,41,42], especially under open-circuit conditions. The atomic displacements in the d1T and Trimer phases, compared with the 1T phase, are on the order of 0.01 Å, as reported in Table S3. Besides, as we report in the Section VII of the SM, the energy gain of forming orbital order in 1T$_3$ phase and the relative stability between d1T and Trimer phases can be tuned by in-plane strain.

Additional studies indicate that the spontaneous polarizations induced by orbital order are not restricted to a specific material but are observed in a group of magnetic transition metal di-halide monolayer compounds as it is illustrated in Fig. 4. The relates results and discussion are reported in the Section VIII of SM. These results confirm that the orbital order in 1T$_3$ phase is the consequence of the interaction between partial occupied t$_{2g}$ or e$_g$ orbitals.

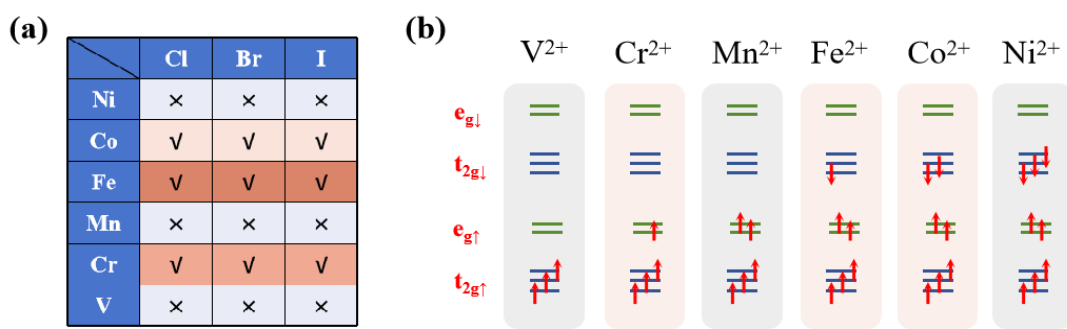

**Fig.4** (a) Heat map reacting to the ground state distribution of $MX_2$. Checkmark and cross indicate whether a 1T to $1T_3$ phase transition has occurred. Darker colors indicate that phase changes are more likely to occur. (b) Orbital occupation of transition metals in the ionic state after bonding with halogen elements.

From the aforementioned results, we can confirm that spontaneous spatial orbital order, breaking the inversion symmetry, acts as the primary driving force, introducing finite OOP plane polarization in magnetic transition metal di-halide monolayer compounds. Two additional points need to be mentioned here. First, similar transition from 1T to d1T phase in other monolayer compounds are reported, such as in $MoS_2$ and $MoTe_2$ [37,39,40]. However, these transitions are primarily dominated by the lattice instability mechanism, the driving force of which is often described by unstable phonon modes rather than electronic origin. The asymmetric orbital order and Jahn-Teller (JT) distortion also appear but act as secondary effects. Distinguished with $MoS_2$ and $MoTe_2$, the spontaneous orbital order and symmetry breaking in $FeI_2$ are independent of lattice distortions. These results also imply the complex competing and cooperating interaction between different order parameters existing in this group of monolayer compounds.

Second, given that the polar Jahn-Teller (JT) distortion is coupled with the orbital order and polarization, it is conceivable that electric field control of JT distortion can be achieved in monolayer compounds of transition metal di-halides (TMDH). Typically, the JT effect is associated with non-polar distortions in most perovskites, making it decoupled from polarization or external electric fields. However, the coupling can establish a connection between JT distortion and polar orders in some perovskites [43]. In the studied TMDH, the symmetry breaking caused by orbital order and the polar group to which the $1T_3$ phase belongs allows for asymmetric JT distortions, making electric control of the JT effect possible. We report our preliminary test in the Fig. S14 of SM.

In summary, we proposed in this work that in magnetic TMDH compounds, spontaneous spatial orbital order introduces two-dimensional ferroelectricity with finite OOP polarization. The highly symmetric 1T phase of $MX_2$ (M=Fe and Cr, X=Cl, Br, I) is not energetically preferred and will spontaneously form spatial orbital order, resulting in significant energy reduction. Due to the inversion symmetry breaking caused by orbital order, the lattice transitions to a polar group with the presence of finite OOP polarization as a secondary effect. We would highlight that this orbital order-induced two-dimensional ferroelectricity present relatively significant larger polarization comparing with those reported typical multiferro materials. Our work reveals an additional pathway for two-dimensional multiferroics and potential applications in electronic and spintronic devices.

**Acknowledgements**

This work was financially supported by the National Natural Science Foundation of China (NSFC, Grant Nos.12274145, 12222402 and 92365101), Guangdong Basic and Applied Basic Research Foundation, China (Grand No. 2023A1515010672). Guangdong Provincial University Science and Technology Program (Grant No. 2023KTSCX029). Work at Fudan is paritally supported by NSFC (grants No. 12188101), Shanghai Science and Technology Program (No. 23JC1400900), the Guangdong Major Project of the Basic and Applied Basic Research (Future functional materials under extreme conditions--2021B0301030005), and Shanghai Pilot Program for Basic Research—FuDan University 21TQ1400100 (23TQ017). X. H acknowledges financial support from F.R.S.-FNRS through the PDR Grants PRO-MOSPAN (T.0107.20), Jin-Zhu Zhao acknowledges the startup funding of South China Normal University.

X-F. L. and X. H. contributed equally to this work.

**References**

[1] M. Bibes, *Nanoferronics Is a Winning Combination*, Nature Mater **11**, 354 (2012).

[2] M. Fiebig, T. Lottermoser, D. Meier, and M. Trassin, *The Evolution of Multiferroics*, Nat Rev Mater **1**, 16046 (2016).

[3] P. Man, L. Huang, J. Zhao, and T. H. Ly, *Ferroic Phases in Two-Dimensional Materials*, Chem. Rev. **123**, 10990 (2023).

[4] C. Xu, H. Yu, J. Wang, and H. Xiang, *First-Principles Approaches to Magnetoelectric Multiferroics*, Annu. Rev. Condens. Matter Phys. 15, 85 (2024).

[5] T. Kimura, T. Goto, H. Shintani, K. Ishizaka, T. Arima, and Y. Tokura, *Magnetic Control of Ferroelectric Polarization*, Nature **426**, 55 (2003).

[6] J. Wang et al., *Epitaxial $BiFeO_3$ Multiferroic Thin Film Heterostructures*, Science **299**, 1719 (2003).

[7] W. Luo, K. Xu, and H. Xiang, *Two-Dimensional Hyperferroelectric Metals: A Different Route to Ferromagnetic-Ferroelectric Multiferroics*, Phys. Rev. B **96**, 235415 (2017).

[8] M. Wu and P. Jena, *The Rise of Two-dimensional van Der Waals Ferroelectrics*, WIREs Comput Mol Sci **8**, e1365 (2018).

[9] C. Huang, Y. Du, H. Wu, H. Xiang, K. Deng, and E. Kan, *Prediction of Intrinsic Ferromagnetic Ferroelectricity in a Transition-Metal Halide Monolayer*, Phys. Rev. Lett. **120**, 147601 (2018).

[10] M. Wu and X. C. Zeng, *Intrinsic Ferroelasticity and/or Multiferroicity in Two-Dimensional Phosphorene and Phosphorene Analogues*, Nano Lett. **16**, 3236 (2016).

[11] W. Li and J. Li, *Ferroelasticity and Domain Physics in Two-Dimensional Transition Metal Dichalcogenide Monolayers*, Nat Commun **7**, 10843 (2016).

[12] H. Wang and X. Qian, *Two-Dimensional Multiferroics in Monolayer Group IV Monochalcogenides*, 2D Mater. **4**, 015042 (2017).

[13] X. Li, L. Li, and M. Wu, *Various Polymorphs of Group III–VI (GaSe, InSe, GaTe) Monolayers with Quasi-Degenerate Energies: Facile Phase Transformations, High-Strain Plastic Deformation, and Ferroelastic Switching*, Materials Today Physics **15**, 100229 (2020).

[14] T. Zhang, Y. Liang, X. Xu, B. Huang, Y. Dai, and Y. Ma, *Ferroelastic-Ferroelectric Multiferroics in a Bilayer Lattice*, Phys. Rev. B **103**, 165420 (2021).

[15] H. J. Xiang, E. J. Kan, Y. Zhang, M.-H. Whangbo, and X. G. Gong, *General Theory for the Ferroelectric Polarization Induced by Spin-Spiral Order*, Phys. Rev. Lett. **107**, 157202 (2011).

[16] J. Y. Ni, X. Y. Li, D. Amoroso, X. He, J. S. Feng, E. J. Kan, S. Picozzi, and H. J. Xiang, *Giant Biquadratic Exchange in 2D Magnets and Its Role in Stabilizing Ferromagnetism of $NiCl_2$ Monolayers*, Phys. Rev. Lett. **127**, 247204 (2021).

[17] Q. Song et al., *Evidence for a Single-Layer van Der Waals Multiferroic*, Nature 602, 601 (2022).

[18] X. Li, C. Xu, B. Liu, X. Li, L. Bellaiche, and H. Xiang, *Realistic Spin Model for Multiferroic $NiI_2$*, Phys. Rev. Lett. **131**, 036701 (2023).

[19] T. Misawa, J. Yoshitake, and Y. Motome, *Charge Order in a Two-Dimensional Kondo Lattice Model*, Phys. Rev. Lett. **110**, 246401 (2013).

[20] C. Wang, X. Zhou, Y. Pan, J. Qiao, X. Kong, C.-C. Kaun, and W. Ji, *Layer and Doping Tunable Ferromagnetic Order in Two-Dimensional $CrS_2$ Layers*, Phys. Rev. B **97**, 245409 (2018).

[21] Y. Zhang, P. Han, M. P. K. Sahoo, X. He, J. Wang, and P. Ghosez, *Designing Orbital and Charge Ordering Multiferroics by Superlattice Strategy and Strain Engineering*, Phys. Rev. B **106**, 235156 (2022).

[22] H.-S. Jin, K.-H. Ahn, M.-C. Jung, and K.-W. Lee, *Strain and Spin-Orbit Coupling Induced Orbital Ordering in the Mott Insulator $BaCrO_3$*, Phys. Rev. B **90**, 205124 (2014).

[23] K. Yamauchi and P. Barone, *Electronic Ferroelectricity Induced by Charge and Orbital Orderings*, J. Phys.: Condens. Matter **26**, 103201 (2014).

[24] A. T. Lee and C. A. Marianetti, *Structural and Metal-Insulator Transitions in Rhenium-Based Double Perovskites via Orbital Ordering*, Phys. Rev. B **97**, 045102 (2018).

[25] K. Zollner, P. E. F. Junior, and J. Fabian, *Strain-Tunable Orbital, Spin-Orbit, and Optical Properties of Monolayer Transition-Metal Dichalcogenides*, Phys. Rev. B **100**, 195126 (2019).

[26] R. Cong, R. Nanguneri, B. Rubenstein, and V. F. Mitrović, *Evidence from First-Principles Calculations for Orbital Ordering in $Ba_2NaOsO_6$: A Mott Insulator with Strong Spin-Orbit Coupling*, Phys. Rev. B **100**, 245141 (2019).

[27] N. A. Hill, *Why Are There so Few Magnetic Ferroelectrics?*, J. Phys. Chem. B **104**, 6694 (2000).

[28] G. Colizzi, A. Filippetti, and V. Fiorentini, *Multiferroicity and orbital ordering in $Pr_{0.5}Ca_{0.5}MnO_3$ from first principles*, Phys. Rev. B 82, 140101 (2010).

[29] K. Gupta, P. Mahadevan, P. Mavropoulos, and M. Ležaić, *Orbital-Ordering-Induced Ferroelectricity in $SrCrO_3$*, Phys. Rev. Lett. 111, 077601 (2013).

[30] K. Singh, C. Simon, E. Cannuccia, M.-B. Lepetit, B. Corraze, E. Janod, and L. Cario, Orbital-Ordering-Driven Multiferroicity and Magnetoelectric Coupling in

GeV$_4$S$_8$, Phys. Rev. Lett. 113, 137602 (2014).

[31] H. Katsura, N. Nagaosa, and A. V. Balatsky, *Spin Current and Magnetoelectric Effect in Noncollinear Magnets*, Phys. Rev. Lett. **95**, 057205 (2005).

[32] J.-J. Zhang, L. Lin, Y. Zhang, M. Wu, B. I. Yakobson, and S. Dong, *Type-II Multiferroic Hf$_2$VC$_2$F$_2$ MXene Monolayer with High Transition Temperature*, J. Am. Chem. Soc. **140**, 9768 (2018).

[33] H. Tan, M. Li, H. Liu, Z. Liu, Y. Li, and W. Duan, *Two-Dimensional Ferromagnetic-Ferroelectric Multiferroics in Violation of the d$_0$ Rule*, Phys. Rev. B **99**, 195434 (2019).

[34] S. Dong, H. Xiang, and E. Dagotto, *Magnetoelectricity in Multiferroics: A Theoretical Perspective*, National Science Review **6**, 629 (2019).

[35] K. Dou, Z. He, W. Du, Y. Dai, B. Huang, and Y. Ma, D$_0$ *Magnetic Skyrmions in Two-Dimensional Lattice*, Adv Funct Materials **33**, 2301817 (2023).

[36] W.-J. Shuai, R. Wang, and J.-Z. Zhao, *Ferroelectric Phase Transition Driven by Anharmonic Lattice Mode Coupling in Two-Dimensional Monolayer In$_2$Se$_3$*, Phys. Rev. B **107**, 155427 (2023).

[37] L.-B. Wan, B. Xu, P. Chen, and J.-Z. Zhao, *Ferroelectric Phase Transition in a 1T Monolayer of MoTe$_2$ : A First-Principles Study*, Phys. Rev. B **108**, 165430 (2023).

[38 X. Cai, Z. Xu, S.-H. Ji, N. Li, and X. Chen, Molecular Beam Epitaxy Growth of Iodide Thin Films, Chinese Phys. B 30, 028102 (2021).

[39] S. N. Shirodkar and U. V. Waghmare, Emergence of Ferroelectricity at a Metal-Semiconductor Transition in a 1T Monolayer of MoS$_2$, Phys. Rev. Lett. 112, 157601 (2014).

[40] S. Yuan, X. Luo, H. L. Chan, C. Xiao, Y. Dai, M. Xie, and J. Hao, *Room-Temperature Ferroelectricity in MoTe$_2$ down to the Atomic Monolayer Limit*, Nat. Commun. **10**, 1775 (2019).

[41] C. Cui et al., Intercorrelated In-Plane and Out-of-Plane Ferroelectricity in Ultrathin Two-Dimensional Layered Semiconductor In$_2$Se$_3$, Nano Lett. 18, 1253 (2018).

[42] J. Xiao et al., Intrinsic Two-Dimensional Ferroelectricity with Dipole Locking, Phys. Rev. Lett. 120, 227601 (2018).

[43] J. Varignon, N. C. Bristowe, and P. Ghosez, *Electric Field Control of Jahn-Teller Distortions in Bulk Perovskites*, Phys. Rev. Lett. **116**, 057602 (2016).